\documentclass[12pt,reqno]{amsart}
\usepackage{mathrsfs}
\usepackage{amsmath}
\usepackage{amssymb, amsmath}

\newcommand{\newcom}{\newcommand}

\newcom{\al}{\alpha}
\newcom{\be}{\beta}
\newcom{\eps}{\epsilon}
\newcom{\ga}{\gamma}
\newcom{\Ga}{\Gamma}
\newcom{\ka}{\kappa}
\newcom{\Lam}{\Lambda}
\newcom{\lam}{\lambda}
\newcom{\Om}{\Omega}
\newcom{\om}{\omega}
\newcom{\Si}{\Sigma}
\newcom{\si}{\sigma}
\newcom{\tht}{\theta}
\newcom{\dtri}{\nabla}
\newcom{\tri}{\triangle}
\newcom{\oo}{\infty}
\newcom{\vphi}{\varphi}
\newcom{\calb}{{\mathcal B}}
\newcom{\calc}{{\mathcal C}}
\newcom{\cald}{{\mathcal D}}
\newcom{\calf}{{\mathcal F}}
\newcom{\calp}{{\mathcal P}}
\newcom{\cals}{{\mathcal S}}
\newcom{\caly}{{\mathcal Y}}
\newcom{\calz}{{\mathcal Z}}
\newcom{\bfz}{{\bf Z}}
\newcom{\R}{\Bbb R}
\newcom{\N}{\Bbb N}
\newcom{\Z}{\Bbb Z}
\newcom{\C}{\Bbb C}
\newcom{\E}{\Bbb E}

\newcom{\f}{\frac}
\newcom{\di}{\displaystyle\int}
\newcom{\ds}{\displaystyle\sum}
\newcom{\dl}{\displaystyle\lim}
\newcom{\ov}{\overline}
\newcom{\sset}{\subset}
\newcom{\wt}{\widetilde}
\newcom{\pp}{\partial}
\newcom{\co}{\cdot}
\newcom{\suml}{\sum\limits}
\newcom{\supl}{\sup\limits}
\newcom{\intl}{\int\limits}
\newcom{\infl}{\inf\limits}
\newcom{\disp}{\displaystyle}
\newcom{\non}{\nonumber}
\newcom{\no}{\noindent}
\newcom{\QED}{$\square$}

\newtheorem{athm}{\bf \t}[section]
\newenvironment{thm} [1] {\def\t{#1}\begin{athm} \bf \rm} {\end {athm}}
\newcom{\bthm}{\begin{thm}}\newcom{\ethm}{\end{thm}}

\newcom{\beq}{\begin{equation}}
\newcom{\eeq}{\end{equation}}
\newcom{\ben}{\begin{eqnarray}}
\newcom{\een}{\end{eqnarray}}
\newcom{\beno}{\begin{eqnarray*}}
\newcom{\eeno}{\end{eqnarray*}}

\numberwithin{equation}{section}

\begin{document}

\title[]
{Long time Anderson localization for nonlinear random Schr\"{o}dinger equation}

\author{W.-M. Wang}
\address{Departement de Mathematique, Universite Paris Sud, 91405 Orsay Cedex, FRANCE} 
\email{wei-min.wang@math.u-psud.fr}

\author{Zhifei Zhang}
\address{School of Mathematical Sciences, Peking University, 100871, P. R. China and
Departement de Mathematique, Universite Paris Sud, 91405 Orsay Cedex, FRANCE }

 \email{zfzhang@math.pku.edu.cn}

\thanks{}

\date{}

\keywords{}

\subjclass[2000]{}


\maketitle

\section{Introduction}

We consider the lattice nonlinear random Schr\"{o}dinger equation in $1-d$:
\beq\label{1.1}
i\dot{q}_j=v_jq_j+\eps_1(q_{j-1}+q_{j+1})+\eps_2|q_j|^2q_j,\quad j\in\Z,
\eeq
where $V=\{v_j\}$ is a family of independent identically distributed (i.i.d.) random variables in $[0,1]$ with uniform distribution, $0<\eps_1,\,\eps_2\ll 1$, the cubic nonlinearity models the particle-particle interaction.  When $\eps_2=0$,
(1.1) is the well studied Anderson tight binding model, where it is known \cite{GMP} that $\forall$ $\epsilon_1$, 
the Schr\" odinger operator
\beq\label{1.2}
H=\epsilon_1\Delta+V\qquad \text{ on }\ell^2(\Bbb Z),\eeq
where 
\beq \label{1.3} \aligned \Delta_{jk}&=1,\quad |i-j|_{\ell^1}=1,\\
&=0,\quad \text{otherwise},\endaligned\eeq
almost surely has pure point spectrum with exponentially localized eigenfunctions. In $d\geq 2$, it is known
\cite{FS, vDK, AM} that for $0<\epsilon_1\ll 1$ almost surely the spectrum is pure point with exponentially localized eigenfunctions. This is called Anderson localization (A.L.) By the RAGE theorem \cite{AG, E, R} (cf. also \cite{CFKS}) pure point spectrum is equivalent to the following statement:

$\forall$ initial datum $\{q_j(0)\}\in\ell^2$, $\delta>0$, $\exists\, j_0$ such that 
\beq\label{1.4}\sup_{t\in\Bbb R}
\sum_{|j|>j_0}|q_j(t)|^2\leq\delta.\eeq
 
When $\epsilon_2\neq 0$, spectral theory is no longer available. However we can still retain (1.4) as a 
criterion for the nonlinear equation (1.1). In this paper, we work in $\ell^2$, the space for the linear theory. 
This is possible as it is easily seen that (1.1) has a global solution in $\ell^2$ and the $\ell^2$ norm of the 
solution $\{q_j(t)\}$ is conserved, i.e., 
\ben
\sum_{j\in \Z}|q_j(t)|^2=\sum_{j\in \Z}|q_j(0)|^2, \quad \forall t\in \R.
\een

Let $\epsilon=\epsilon_1+\epsilon_2$. We prove

\bthm{Theorem} Given $\delta>0, A>1$, for all initial datum $\{q_j(0)\}_{j\in \Z}\in\ell^2$,
let $j_0\in\Bbb N$ be such that 
\beq
\sum_{|j|> j_0}|q_j(0)|^2<\delta.
\eeq
Then there exist $C=C(A)>0$,  $\epsilon(A)>0$ and $N=N(A)>A^2$ such that for all $ t\leq ({\delta /C})\eps^{-A},$
\beq
\sum_{|j|>j_0+N}|q_j(t)|^2<2\delta,
\eeq
with probability
\beq
1-e^{-\f {j_0} {N}e^{-2N\eps^{\f1{CA}}}},
\eeq
provided $0<\epsilon<\epsilon(A)$.
\ethm

We interpret the above result as long time Anderson localization for the nonlinear random Schr\"odinger
equation (1.1). The proof uses Birkoff normal form type transformations. The main feature of this normal form
is that it contains energy barriers centered at some $\pm j_0\in\Bbb Z$, $j_0>1$ of width $N$, where the terms 
responsible for mode propagation are small $\sim\epsilon^A$. This is similar to the normal form transform in
\cite{BW1}. The fact that the transformation is only in a small neighborhood enables us to treat 
$\ell^2$ data. They are ``rough" data when viewing $j\in\Bbb Z$ as a Fourier index.

This normal form is different from the usual Birkoff normal form used in nonlinear PDE's, cf. e.g., \cite{BG}, where one 
typically needs smooth initial data, which in the present context means that $\{q_j(0)\}$ such that 
$\sum j^{2s}|q_j(0)|^2=1$ for $s>s_0\gg 1$. The present method seems particularly suited to treat
nonlinear lattice Schr\" odinger equations, where typically one has short range interactions.

We now comment on a fine point, namely the small parameter $\epsilon_1$ in (1.1), which was not needed
in $1-d$ to prove A.L. for the linear equation. The reason we need it for the nonlinear equation is because
we need to exclude certain potential configurations in addition to what is needed for A.L. This is in order
to avoid small denominators which correspond to new resonances generated by the nonlinearity. Since 
this exclusion is {\it a postiori}, had we used the bases provided by the eigenfunctions we would have needed precise information on how the eigenfunctions vary as the potential varies. To our knowledge, this does not seem 
to be available in the existing literature.  

The above theorem raises the natural question of the limit as $t\to\infty$ (independent of $\epsilon_1$ and $\epsilon_2$). In
\cite{BW2}, time quasi-periodic solutions were constructed in all dimensions for small $\epsilon_1$ and $\epsilon_2$. (Previously a special type of time periodic solutions where there is only the basic frequency was constructed in \cite{AF}.)
Certainly in that case (1.4) remains valid as $t\to\infty$. The validity or invalidity of (1.4) as $t\to\infty$
for more general initial data remains essentially an open problem.

W.-M. Wang thanks M. Aizenman for a stimulating lecture in Paris, which motivated this paper. She also 
thanks T. Spencer for initiation to the subject of A.L. and for numerous enlightening discussions. 

\section{Structure of transformed Hamiltonian}

We recast (\ref{1.1}) as a Hamiltonian equation:
\ben
i\dot{q}_j=2\f {\pp H} {\pp \overline{q}_j},\label{2.1}
\een
with the Hamiltonian
\ben
H(q,\bar q)=\f12\bigl(\sum_{j\in \Z}v_j|q_j|^2+\eps_1\sum_{j\in \Z}(\bar q_jq_{j+1}+q_j\bar q_{j+1})
+\f12\eps_2\sum_{j\in \Z}|q_j|^4\bigr).\label{2.2}
\een
As mentioned earlier,  the $\ell^2$ norm of the solution $\{q_j(t)\}$ is conserved, i.e.
\ben
\sum_{j\in \Z}|q_j(t)|^2=\sum_{j\in \Z}|q_j(0)|^2, \quad \forall t\in \R.\label{2.3}
\een

In order to prove (1.7), we need to control the time derivative of the truncated sum of higher modes
\beq
\f {d} {dt}\sum_{|j|>j_0}|q_j(t)|^2.\label{2.4}
\eeq
As in \cite{BW1}, we will use the random potential $V=\{v_j\}_{j\in \Z}$ to obstruct energy
transfer from low to high modes by creating ``zones'' in $\Z$, where the only mode coupling term is of order $O(\eps^A)$,
where $\epsilon=\epsilon_1+\epsilon_2$ as before. This construction is achieved by invoking the usual process of symplectic transformations.

In what follows, we will deal extensively with monomials in $q_j$. So we first introduce some notations. Rewrite any
monomials in the form:
$$
\prod_{j\in \Z}q_j^{n_j}\bar q_j^{n_j'}.
$$
Let $n=\{n_j,n_j'\}_{j\in \Z}\in \N^{\Z}\times \N^{\Z}$. We will use three notations: support, diameter and degree:
\beno
\textrm{supp}\, n&=&\{j|n_j\neq 0\, \textrm{or}\, n_j'\neq 0\}\\
\Delta(n)&=& \textrm{diam}\{\textrm{supp}\, n\}\\
|n|&=& \sum_{j\in \textrm{supp}\, n}(n_j+n_j').
\eeno
If $n_j=n_j'$ for all $j\in \textrm{supp}\, n$, then the monomial is called resonant.
Otherwise it is called non-resonant. Note that non-resonant monomials contribute to the truncated
sum in (\ref{2.4}), while resonant ones do not.

To control the sum in (\ref{2.4}), we will transform  $H$ in (\ref{2.2}) to $H'$ of the form:
\ben
H' &=&\f12\sum_{j\in \Z}(v_j+w_j)|q_j|^2\label{2.5}\\
&&+\sum_{n\in \N^{\Z}\times \N^{\Z}}c(n)\prod_{\textrm{supp}\,n}q_j^{n_j}\bar q_j^{n_j'}\label{2.6}\\
&&+\sum_{n\in \N^{\Z}\times \N^{\Z}}d(n)\prod_{\textrm{supp}\,n}|q_j|^{2n_j}\label{2.7}\\
&&+O(\eps^A),\label{2.8}
\een
where (\ref{2.6}) consists of non-resonant monomials ($n_j\neq n_j'$ for some $j$),
(\ref{2.7}) consists of resonant monomials of degree at least 4. Note that
\ben
 \sum_{\textrm{supp}\, n} n_j=\sum_{\textrm{supp}\, n} n_j'\label{2.9}
\een
for  non-resonant monomials in (\ref{2.6}), which is a general feature of polynomial Hamiltonian.
The coefficients $c(n)$ and $d(n)$ satisfy the bound
\ben
\forall n,\quad |c(n)|+|d(n)|<\exp(-\rho\{\Delta(n)+|n|-2\}\log{\f 1\eps}),\quad \rho>\f 1 {10}.\label{2.10}
\een

The transformed Hamiltonian $H'$ will manifest an ``energy barrier ''. More precisely,
we require that
\ben
|c(n)|<\eps^A, \quad \textrm{if}\,\, \textrm{supp}\, n\cap \{[-b,-a]\cup [a,b]\}\neq \emptyset,\label{2.11}
\een
where $a$ and $b$ satisfy
\ben
[j_0-\f N2,j_0+\f N2]\subset [a,b] \subset [j_0-N,j_0+N],\label{2.12}
\een
and $N$ is an integer depending on $A$.

In (\ref{2.5}), $w_j$ and all coefficients $c(n),d(n)$ depend on $V$. Let $W=\{w_j\}_{j\in \Bbb Z}$.
We require that
\ben
|\nabla_Vc(n)|+|\nabla_Vd(n)|<1\label{2.13}
\een
and
\ben
\|\f {\partial W} {\partial V}\|_{\ell^2\rightarrow \ell^2}<\eps^{\f1{40}}.\label{2.14}
\een

The transformation from $H$ to $H'$ will be achieved by a finite step iterative process. Let $H_s, \Gamma_s$
be the Hamiltonian and the transformation at step $s$, $H_{s+1}=H_s\circ \Ga_s$. At each step $s, \Ga_s$ is
the symplectic transformation generated by an appropriate polynomials Hamiltonian $F$.
$H_{s+1}$ is the time-1 map, computed by using a convergent Taylor series of successive Poisson brackets of
$H_s$ and $F$, i.e.
\beno
H_{s+1}=H_s\circ \Ga_s=\{H_s,F\}+\f 1 {2!}\{\{H_s,F\},F\}+\cdots,
\eeno
where the Poisson bracket $\{H_s,F\}$ is defined by
\beno
\{H_s,F\}=\sum_{j\in \Z}\f {\partial H_s} {\partial \bar q_j}\f {\partial F} {\partial q_j}-
\f {\partial H_s} {\partial q_j}\f {\partial F} {\partial \bar q_j}.
\eeno

It is  important to remark that  our construction only involves the modes $j\in \Z$ for which
$||j|-j_0|\le N$. So, if
$$\textrm{supp}\, n\cap \{[-j_0-N,-j_0+N]\cup [j_0-N,j_0+N]\}\neq \emptyset$$
in the sum of (\ref{2.6}-\ref{2.7}), then we have by the fact that $\Delta(n)\le 1$ for all the terms in $H$ that
$$
\textrm{supp}\, n\subset [-j_0-N-1,-j_0+N+1]\cup [j_0-N-1,j_0+N+1],
$$
which together with (\ref{2.10}) implies that we can assume that
\ben
\Delta(n)<20A\quad \textrm{and} \quad |n|<20A\label{2.15}
\een
for the terms with
$\textrm{supp}\, n\cap \{[-j_0-N,-j_0+N]\cup [j_0-N,j_0+N]\}\neq \emptyset$,
the other terms can be captured by (\ref{2.8}) in $H'$. On the other hand, one has 
\ben
|c(n)|,\,|d(n)|\leq \eps, \quad \Delta(n)\le 1 \label{2.16}
\een
for the terms with $\textrm{supp}\, n\cap \{[-j_0-N,-j_0+N]\cup [j_0-N,j_0+N]\}=\emptyset$.

In addition, $w_j=0$ unless $||j|-j_0|\le N+1$, which together with (\ref{2.14}) implies that the frequency
modulation map $V\rightarrow \widetilde{V}=V+W$ satisfies
\ben
e^{-2N\eps^{\f1{40}}}<(1-\eps^{\f1{40}})^{2N+2}<\bigl|\textrm{det}\f {\partial {\widetilde{V}}} {\partial V}\bigr|
<(1+\eps^{\f1{40}})^{2N+2}<e^{2N\eps^{\f1{40}}}.\label{2.17}
\een
The non-resonance estimates in section 3 on symplectic transforms are expressed in terms of
$\widetilde{V}$. These non-resonance estimates will be translated into probabilistic estimates in $V$
by using (\ref{2.17}).

\section{Analysis and estimates of the symplectic transformations}

We now construct the symplectic transformation $\Gamma$ so that the transformed Hamiltonian $H'=H\circ\Gamma$ satisfies
(\ref{2.10})-(\ref{2.14}). It is achieved by a finite step induction. At the first step: $s=1$
\beno
H_1=H=\f12\bigl(\sum_{j\in \Z}v_j|q_j|^2+\eps_1\sum_{j\in \Z}(\bar q_jq_{j+1}+q_j\bar q_{j+1})
+\f12\eps_2\sum_{j\in \Z}|q_j|^4\bigr).
\eeno
Let $\eta_j$ denote the canonical basis of $\Z$, (\ref{2.10}-\ref{2.12}) are satisfied with
\beno
c(n)=c(\eta_j\times \eta_{j+1})&\leq&\f \eps 2,\quad |n|=2, \Delta(n)=1, \textrm{supp}\,n=\{j,j+1\},\\
&=& 0\quad \textrm{otherwise}.\\
d(n)=d(\eta_j\times \eta_{j+1})&\leq&\f \eps 4,\quad |n|=4, \Delta(n)=0, \textrm{supp}\,n=\{j\},\\
&=& 0\quad \textrm{otherwise}.
\eeno
(\ref{2.13}) is trivially satisfied with
\beno
|\nabla_V(c)|=|\nabla_V(d)|=0,
\eeno
and so is (\ref{2.14}):
\beno
W=0,\qquad \f {\partial W} {\partial V}=0.
\eeno

Assume that we have obtained at step $s$, the Hamiltonian $H_s$ in the form (\ref{2.5}-\ref{2.8}) satisfying
$(\ref{2.10})_s-(\ref{2.14})_s$. Our aim is to produce $H_{s+1}$ possessing the corresponding properties at step $s+1$.
In what follows, $\eps$ always denotes a sufficiently small constant depending only on $A$.
$(\ref{2.10})_s-(\ref{2.12})_s$ state that
\ben
|c(n)|+|d(n)|<\exp(-\rho_s\{\Delta(n)+|n|-2\}\log{\f 1\eps}),\label{3.1}
\een
with $\rho_s>\f 1 {10}$, moreover
\ben
|c(n)|<\delta_s \quad \textrm{if}\,\,\, \textrm{supp}\, n\cap \{[-b_s,-a_s]\cup [a_s,b_s]\}\neq \emptyset,\label{3.2}
\een
with
\ben
[j_0-\f N2,j_0+\f N2]\subset [a_s,b_s] \subset [j_0-N,j_0+N],\label{3.3}
\een
and $\delta_s$ is defined inductively as
\ben
\delta_1=\f \eps 2,\quad \delta_s=\delta_{s-1}^{\f {19} {10}}+\eps^{\f 1 {20}}\delta_{s-1},\quad s\ge 2.\label{3.4}
\een
We remark that the first term in (\ref{3.4}) comes from Poisson brackets of polynomials with coefficients $c$, while the
second one comes from Poisson brackets of polynomials with coefficients $c$ and $d$.

We satisfy (\ref{3.2}) at step $s+1$ constructively by removing those $c(n)$ with $\delta_{s+1}<|c(n)|<\delta_s$,
$\textrm{supp}\, n\cap \{[-b_s,-a_s]\cup [a_s,b_s]\}\neq \emptyset$ and a corresponding reduction of
$[-b_s,-a_s]\cup [a_s,b_s]$ to $[-b_{s+1},-a_{s+1}]\cup [a_{s+1},b_{s+1}]$ with $a_{s+1}>a_s, b_{s+1}<b_s$,
so that
\ben
|c(n)|\le \delta_{s+1} \quad
\textrm{if}\quad \textrm{supp}\, n\cap \{[-b_{s+1},-a_{s+1}]\cup [a_{s+1},b_{s+1}]\}\neq \emptyset.\label{3.5}
\een

We proceed as follows. Denoting in $H_s$ (\ref{2.5})-(\ref{2.8}),
\ben
\widetilde{v}_j=v_j+w_j^{(s)}\quad \textrm{and }\quad H_0=\sum_{j\in \Z}\widetilde{v}_j|q_j|^2,\label{3.6}
\een
we define, following the standard approach
\ben
H_{s+1}=H_s\circ \Gamma_F,\label{3.7}
\een
where $\Gamma_F$ is the symplectic transformation obtained from the Hamiltonian function
\beno
F=\sum_{\textrm{supp}\, n\subset [-b_s,-a_s]\cup [a_s,b_s],|c(n)|>\delta_{s+1}}\f {c(n)}
{\sum (n_j-n_j')\widetilde{v}_j}\prod q_j^{n_j}\bar q_j^{n_j'}.
\eeno
Here we need to impose the small divisor condition
\ben
|\sum (n_j-n_j')\widetilde{v}_j|>\delta_s^{\f 1 {100s^2}},\label{3.8}
\een
which will lead to measure estimates of this construction in section 4.

Recall that $H_{s+1}$ is the time-1 map and by Taylor series:
\ben
H_{s+1}=H_s\circ \Gamma_F&=&H_0+(\ref{2.6})+\{H_0,F\}+(\ref{2.7})\nonumber\\
&&+\{(\ref{2.6}),F\}+\f 1 {2!}\{\{(\ref{2.6}),F\},F\}+\cdots\label{3.9}\\
&&+\{(\ref{2.7}),F\}+\f 1 {2!}\{\{(\ref{2.7}),F\},F\}+\cdots\label{3.10}\\
&&+(\ref{2.8})\circ \Gamma_F.\nonumber
\een
Note that
\beno
\{H_0,F\}=\sum_{\textrm{supp}\, n\subset [-b_s,-a_s]\cup [a_s,b_s],|c(n)|>\delta_{s+1}}c(n)
\prod q_j^{n_j}\bar q_j^{n_j'},
\eeno
and if $|c(n)|>\delta_{s+1}$, then by (\ref{3.1})
\beno
\Delta(n)< 10\f {\log\f1{\delta_{s+1}}} {\log\f1\eps}.
\eeno
Define
\ben\label{3.11}
a_{s+1}=a_s+20\f {\log\f1{\delta_{s+1}}} {\log\f1\eps},\quad b_{s+1}=b_s-20\f {\log\f1{\delta_{s+1}}} {\log\f1\eps}.
\een
Then $\{H_0,F\}$ removes in (\ref{2.6}) all monomials for which $|c(n)|>\delta_{s+1}$
and $\textrm{supp}\, n\cap \{[-b_{s+1},-a_{s+1}]\cup [a_{s+1},b_{s+1}]\}\neq \emptyset$.
Thus, $(\ref{2.6})+\{H_0,F\}$ satisfies (\ref{3.5}). Note that $a_{s+1}$ and $b_{s+1}$ do not shrink to $j_0$
for $N$ large enough depending only on $A$( cf. (\ref{3.25})).  We next prove that (\ref{3.9},\ref{3.10})
satisfy $(\ref{3.1},\ref{3.2})_{s+1}$.

\vspace{0.2cm}
\no \textbf{Monomials in (\ref{3.9})}\vspace{0.2cm}

We begin with the first two Poisson brackets. We rewrite them as
\beno
&&\{(\ref{2.6}),F\}=\sum_{\mu}g_1(\mu)\prod q_j^{\mu_j}\bar q_j^{\mu_j'},\\
&&\f 1 {2!}\{\{(\ref{2.6}),F\},F\}=\sum_{\mu}g_2(\mu)\prod q_j^{\mu_j}\bar q_j^{\mu_j'}.
\eeno
The Poisson bracket $\{(\ref{2.6}),F\}$ produces monomials of the form
\ben\label{3.12}
&&\{\prod q_j^{m_j}\bar q_j^{m_j'},\prod q_j^{n_j}\bar q_j^{n_j'}\}\nonumber\\
&&=\sum(m_kn_k'-m_k'n_k)q_k^{m_k+n_k-1}\bar q_k^{m_k'+n_k'-1}\prod_{j\neq k}q_j^{m_j+n_j}\bar q_j^{m_j'+n_j'},
\een
with the coefficient
\beno
\f {c(m)c(n)} {\sum(n_j-n_j')\widetilde{v}_j}
\eeno
where
\beno
\textrm{supp}\, n\subset [a_s,b_s]\cup [-b_s,-a_s],\quad \textrm{supp}\, m\cap \textrm{supp}\, n\neq\emptyset.
\eeno
Hence $\textrm{supp}\, m\cap \{[a_s,b_s]\cup [-b_s,-a_s]\}\neq\emptyset$. Then it follows from (\ref{3.2}) that
\ben\label{3.13}
|c(m)|, |c(n)|<\delta_s.
\een

The monomials in (\ref{3.12}) corresponding to multi-index $\mu$ satisfy
\ben\label{3.14}
\Delta(\mu)\le \Delta(m)+\Delta(n),\quad |\mu|=|m|+|n|-2.
\een
The number of realizations of a fixed monomials $\prod q_j^{\mu_j}\bar q_j^{\mu_j'}$ is bounded by
\ben\label{3.15}
2^{|\mu|}(\Delta(m)\wedge\Delta(n))<2^{|\mu|}(\Delta(m)+\Delta(n)).
\een

Summing up (\ref{3.13}-\ref{3.15}), we get by using the small divisor bound (\ref{3.8}) that
\ben
|g_1(\mu)|\le 2^{|\mu|}(|\mu|+2)^2\delta_s^{-\f 1{100s^2}}|c(m)||c(n)|(\Delta(m)+\Delta(n)).\label{3.16}
\een
Define
\beno
\rho_{s+1}=\rho_s(1-\f 1 {10s^2}).\label{3.17}
\eeno
Then we get by using (\ref{3.1}), (\ref{3.13}) and (\ref{3.14}) that
\ben
|g_1(\mu)|&\le& 2^{|\mu|}(|\mu|+2)^2|\log\delta_s^2|\delta_s^{-\f 1{100s^2}}\delta_s^{\f2 {10s^2}}\nonumber\\
&& \exp(-\rho_{s+1}\{\Delta(\mu)+|\mu|-2\}\log{\f 1\eps}),\label{3.17}
\een
where we used (\ref{3.1}) to bound $\Delta(m),\Delta(n)$ in terms of $c(m), c(n)$.
Note that by (\ref{2.15}) and (\ref{3.14})
\beno
|\mu|<40A,
\eeno
and by (\ref{3.4})
\ben
\log \delta_{s+1}^{-1}\approx s\log\f 1 \eps\label{3.18}
\een
which implies that we can terminate the construction at step $s_*\sim A$ such that $\delta_{s_*}<\epsilon^A$.
Thus (\ref{3.17}) gives that
\ben
g_1(\mu)<\delta_{s}^{\f1{20s^2}}\exp(-\rho_{s+1}\{\Delta(\mu)+|\mu|-2\}\log{\f 1\eps}),\label{3.19}
\een
for $\epsilon$ sufficiently small depending only on $A$.

We now turn to the estimate of $g_2(\mu)$. A fixed monomial $\prod q_j^{\mu_j}\bar q_j^{\mu_j'}$
in $\{\{(\ref{2.6}),F\},F\}$ is now the confluence of 3 sources, denoted by $m,n,p$ with
$$
|\mu|=|m|+|n|+|p|-4.
$$
Continuing the previous terminology, the coefficient is
$$
c(m)c(n)c(p)\qquad \textrm{with }\,\,|c(m)|, |c(n)|, |c(p)|<\delta_s,
$$
and the prefactor is a sum of terms of the form
\beno
&&(m_kn_k'-m_k'n_k)[(m_j+n_j)p_{\ell}'-(m_j'+n_j')p_{\ell}]\quad \textrm{if}\,\, j\neq k \quad \textrm{or}\\
&&(m_kn_k'-m_k'n_k)[(m_k+n_k-1)p_{\ell}'-(m_k'+n_k'-1)p_{\ell}].
\eeno
Hence the prefactor is bounded by $(|\mu|+4)^4$. The entropy is bounded by
$$
\sum_{|w|=2}^{|\mu|}2^{|w|}(\Delta(m)\wedge \Delta(n))2^{|\mu|}(\Delta(w)\wedge \Delta(p))
\le [2^{|\mu|}(\Delta(m)+\Delta(n)+\Delta(p))]^2.
$$
Therefore we have
\ben\label{3.20}
g_2(\mu)&\le& \f 1 {2!}(\delta_s^{-\f 1{100s^2}})^2(|\mu|+4)^4[2^{|\mu|}(\Delta(m)+\Delta(n)+\Delta(p))]^2\delta_s^{\f3 {10s^2}}\nonumber\\
&& \exp(-\rho_{s+1}\{\Delta(\mu)+|\mu|-2\}\log{\f 1\eps})\nonumber\\
&\le&  \f 1 {2!}(\delta_s^{-\f 1{100s^2}})^2[2^{|\mu|}(|\mu|+4)^2]^2|\log\delta_s^3|^2
\delta_s^{\f3 {10s^2}}\nonumber\\
&& \exp(-\rho_{s+1}\{\Delta(\mu)+|\mu|-2\}\log{\f 1\eps})\nonumber\\
&<& (\delta_{s}^{\f1{20s^2}})^2 \exp(-\rho_{s+1}\{\Delta(\mu)+|\mu|-2\}\log{\f 1\eps}).
\een

From (\ref{3.19}, \ref{3.20}), the structure of the estimates on the Poisson brackets in (\ref{3.9}) is clear and we obtain that
the $\prod q_j^{\mu_j}\bar q_j^{\mu_j'}$ factor in (\ref{3.9}) is bounded by
\ben
g(\mu)<\exp(-\rho_{s+1}\{\Delta(\mu)+|\mu|-2\}\log{\f 1\eps}).\label{3.21}
\een
Furthermore, we also have
\ben
g(\mu)<\delta_s^{2-\f 1 {50s^2}}<\delta_{s}^{\f {19} {10}}.\label{3.22}
\een

\vspace{0.2cm}
\no \textbf{Monomials in (\ref{3.10})}\vspace{0.2cm}

We rewrite (\ref{3.10}) as
\beno
(\ref{3.10})=\sum_{\mu}\gamma(\mu)\prod q_j^{\mu_j}\bar q_j^{\mu_j'}.
\eeno
Similar to the proof of (\ref{3.21}) and (\ref{3.22}), we get by using the fact that $|d(n)|\le \eps^\f1{10}$ that
\ben
&&\gamma(\mu)<\delta_{s}^{\f1{20s^2}}\exp(-\rho_{s+1}\{\Delta(\mu)+|\mu|-2\}\log{\f 1\eps}),\label{3.23}\\
&&\gamma(\mu)<\eps^\f1{20}\delta_{s}.\label{3.24}
\een

Summing up (\ref{3.21})-(\ref{3.24}), we conclude that $H_{s+1}$ satisfies $(\ref{3.1},\ref{3.2})_{s+1}$.
We now check $(\ref{3.3})_{s+1}$ for the interval $[a_s,b_s]$. We get by (\ref{3.11}) that
\beno
|a_s-a_{s+1}|+|b_s-b_{s+1}|\le 20\f {\log\f1{\delta_{s+1}}} {\log\f1\eps}.
\eeno
Let $s_*\thicksim A$ be such that $\delta_{s_*}<\eps^A$. Then we have
\ben\label{3.25}
|a_1-a_{s_*}|+|b_1-b_{s_*}|\le \f {20} {\log\f1\eps}\sum_{t\le s_*}\log\f1{\delta_{t}}\lesssim A^2,
\een
and $(\ref{3.3})_{s+1}$ will hold from $a_1=j_0-N, b_1=j_0+N$, if $N\gg A^2$.

Finally, let us check $(\ref{2.13},\ref{2.14})_{s+1}$ for $H_{s+1}$. In  $H_0$, we need to add resonant quadratic terms
produced in (\ref{3.9},\ref{3.10}). Denoting these terms by $w_j^{(s)}, \widetilde{v}_j$ is then perturbed to
\beno
\widetilde{\widetilde{v}}_j=\widetilde{v}_j+w_j^{(s)},
\eeno
where $w_j^{(s)}$ by construction satisfies
\beq\label{3.26}
|w_j^{(s)}|<\delta_{s+1}.
\eeq
Therefore, all non-resonance conditions imposed so far can be replaced by
\ben\label{3.27}
|\sum (n_j-n_j')\widetilde{\widetilde{v}}_j|>\delta_s^{\f 1 {100s^2}},\label{3.8}
\een
for all $t\le s$ and $n$ satisfying
\beno
\textrm{supp}\, n\subset [a_t,b_t]\cup [-b_t,-a_t],\quad \Delta(n)< 10\f {\log\f1{\delta_{s+1}}} {\log\f1\eps}.
\eeno

Finally, we check the $V$ dependence for $g(\mu), \gamma(\mu)$. We have
\ben
|\nabla_Vg_1(\mu)|+|\nabla_V\gamma_1(\mu)|&<&2^{|\mu|}(|\mu|+2)^2(\Delta(m)+\Delta(n))\nonumber\\
&&|\nabla_V\f {c(m)c(n)} {\sum(n_j-n_j')\widetilde{v}_j}|.\label{3.28}
\een
Using (\ref{2.13}) and (\ref{2.14}), we have
\beno
&&|\nabla_V\f {c(m)c(n)} {\sum(n_j-n_j')\widetilde{v}_j}|\\&&\le [(|\nabla_Vc(m)|+|\nabla_Vd(m)|)|c(n)|
+(|c(m)|+|d(m)|)|\nabla_Vc(n)|]|\sum(n_j-n_j')\widetilde{v}_j|^{-1}\\
&&\quad+2N^\f12|n|(|c(m)|+|d(m)|)|c(n)||\sum(n_j-n_j')\widetilde{v}_j|^{-2}\|D \widetilde{V}\|_{\ell^2\rightarrow \ell^2}\\
&&<2\delta_s^{-\f 1{100s^2}}(\delta_s+\epsilon^\f 1 {10}),
\eeno
which together with (\ref{3.27}) gives that
\beno
|\nabla_Vg_1(\mu)|+|\nabla_V\gamma_1(\mu)|<\delta_s^\f23+\epsilon^\f 1{15}.
\eeno
The higher order brackets can be treated similarly and we obtain
\beno
|\nabla_Vg(\mu)|+|\nabla_V\gamma(\mu)|<\delta_s^\f12+\epsilon^\f 1{20}.
\eeno
In particular, $(\ref{2.13})_{s+1}$ holds and morover from Schur's lemma
\beno
\|\f {\partial W^{(s)}} {\partial V}\|_{\ell^2\rightarrow\ell^2}\lesssim (\delta_s^{\f12}+\eps^{\f1{20}})\f {\log\f1{\delta_{s+1}}} {\log\f1\eps}
<\delta_s^{\f13}+\eps^{\f1{30}}
\eeno
as $\Delta(n)<10\f {\log\f1{\delta_{s+1}}} {\log\f1\eps}$
and $\log \delta_{s+1}^{-1}\approx s\log\f 1 \eps,s\le s_*\sim A$. Since $W=\sum_{s=1}^{s_*}W^{(s)}$,
$(\ref{2.14})_{s+1}$ remains valid along the process.

\section{Estimates on measure}

Recall that the estimates on the symplectic transformations in section 3 depend on the non-resonance condition
\ben
|\sum (n_j-n_j')\widetilde{v}_j|>\delta_s^{\f 1 {100s^2}},\label{4.1}
\een
where $n$ satisfies
\ben\label{4.2}
\textrm{supp}\, n\subset [a_s,b_s]\cup [-b_s,-a_s],\quad \Delta(n)< 10\f {\log\f1{\delta_{s+1}}} {\log\f1\eps},\quad |n|\le 20A,
\een
Moreover, $\widetilde{v}_j$ can be replaced by $\widetilde{v}_j^{s_*}$ at last step $s_*$ in view of (\ref{3.26}).
This is convenient as we only need to work with a fixed $\widetilde{v}_j$, namely $\widetilde{v}_j=\widetilde{v}_j^{s_*}$.

We first make measure estimates in $\widetilde{V}$ via (\ref{4.1}), and then convert the estimates to estimates in $V$
by using the Jacobian estimates (\ref{2.17}). Denote for a given $n$
\beno
j_+(n)=\max\{j\in \Z|n_j-n_j'\neq 0\}.
\eeno
The set of acceptable $\widetilde{V}$ contains
\ben\label{4.3}
S=\bigcap_{||k|-j_0|<N}\bigcap_{s=1,\cdots,s_*}\bigcap_{n\, satisfies (\ref{4.2}), j_+(n)=k}
\{\widetilde{V}||\sum_{j\le k} (n_j-n_j')\widetilde{v}_j|>\delta_s^{\f 1 {100s^2}}\}.
\een
Let
\beno
S_k=\bigcap_{s=1,\cdots,s_*}\bigcap_{n\, satisfies (\ref{4.2}), j_+(n)=k}
\{\widetilde{V}||\sum_{j\le k} (n_j-n_j')\widetilde{v}_j|>\delta_s^{\f 1 {100s^2}}\}.
\eeno
Then for fixed $(\widetilde{v}_j)_{j<k}$ (Note strict inequality here.) and $n$ such that $j_+(n)=k$
\beno
\textrm{mes}_{\widetilde{v}_k}\{\widetilde{V}||\sum_{j\le k} (n_j-n_j')\widetilde{v}_j|<\delta_s^{\f 1 {100s^2}}\}<2\delta_s^{\f 1 {100s^2}}.
\eeno
Let $S_k^c$ be the complement of the set $S_k$. Its measure can be estimated as
\ben\label{4.4}
\textrm{mes}S_k^c&\le& \sum_{s=1}^{s_*}\sum_{n\, satisfies (\ref{4.2}), j_+(n)=k}\textrm{mes}_{\widetilde{v}_k}\{\widetilde{V}||\sum_{j\le k} (n_j-n_j')\widetilde{v}_j|<\delta_s^{\f 1 {100s^2}}\}\nonumber\\
&\le& 2\sum_{s=1}^{s_*} (10\f {\log\f1{\delta_{s+1}}} {\log\f1\eps})^{20A}\delta_s^{\f 1 {100s^2}}.
\een
Note that $\log \delta_s^{-1}\sim s\log \f 1\eps, s_*\sim A$. Then (\ref{4.4}) gives that for some positive constant $C$
\beno
\textrm{mes}S_k^c<\eps^{\f1{CA}}.
\eeno
Thus, we get
\beno
\textrm{mes}_{\widetilde{V}}S>(1-\eps^{\f1{CA}})^{2N}>e^{-2N\eps^{\f1{CA}}},
\eeno
which together with the Jacobian estimates gives that
\beno
\textrm{mes}_{V}S>e^{-N\eps^\f1{40}}e^{-2N\eps^{\f1{CA}}}>e^{-3N\eps^{\f1{CA}}}.
\eeno
For fixed $j_0, S$ corresponds to a rare event. To circumvent it, as in \cite{BW1}, we allows $j_0$ to vary
in some interval $[\bar j_0,2\bar j_0]$. Taking into account that the restriction in (\ref{4.1}) only relates to
$v_j|_{||j|-j_0|<N}$, we get by using independence that with probability at least
\ben\label{4.5}
1-(1-e^{-3N\eps^{\f1{CA}}})^{\f {j_0} {2N}}>1-e^{-\f {j_0} {2N}e^{-3N\eps^{\f1{CA}}}}
\een
the condition (\ref{4.1}) holds for some $j_0\in [\bar j_0,2\bar j_0]$, where $\f {j_0} {2N}$ is 
the number of independent interval of length $2N$ in $[\bar j_0,2\bar j_0]$.

\section{Proof of Theorem 1.1}
 In sections 3 and 4, we showed that for fixed $\bar j_0\in \Z$ large enough, there exists a $j_0\in [\bar j_0,2\bar j_0]$ such that with probability
 \beno
1-e^{-\f {j_0} {2N}e^{-3N\eps^{\f1{CA}}}}
 \eeno
$H$ is symplectically  transformed into $H'$:
\begin{equation} \label{5.1}
\begin{aligned}
H'=&\f12\sum_{j\in \Z}\widetilde{v}_j|q_j|^2\\
&+\sum_{n\in \N^{\Z}\times \N^{\Z}}c(n)\prod_{\textrm{supp}\,n}q_j^{n_j}\bar q_j^{n_j'}\\
&+\sum_{n\in \N^{\Z}\times \N^{\Z}}d(n)\prod_{\textrm{supp}\,n}|q_j|^{2n_j}\\
&+O(\eps^A),
\end{aligned}
\end{equation}
where $\widetilde{v}_j$ are the modulated frequencies, $c(n)$ are the coefficients of non-resonant monomials, and
$d(n)$ are the coefficients of resonant monomials. The coefficients $c(n)$ satisfy
\ben\label{5.2}
|c(n)|<\eps^A, \quad \textrm{if}
\een
\beno
\textrm{supp}\, n\cap \{[-j_0-\f N2,-j_0+\f N2]\cup [j_0-\f N2,j_0+\f N2]\}\neq \emptyset.
\eeno

Now we are in a position to complete the proof of Theorem 1.1.
The coordinates $q(t)=\{q_j(t)\}_{j\in \Z}$ satisfy
\ben
i\dot q=\f {\partial H} {\partial \bar q}.\label{5.3}
\een
Denote the new coordinates in $H'$ by $q'$. Then (\ref{5.3}) becomes
\ben
i\dot q'=\f {\partial H'} {\partial \bar q'}.\label{5.4}
\een
We get by using (\ref{5.1}) and (\ref{5.4}) that
\ben\label{5.5}
&&\f {d} {dt} \sum_{|j|>j_0}|q'_j(t)|^2=4\textrm{Im}\sum_{|j|>j_0}\bar q_j'(t)\f {\partial H'} {\partial \bar q_j'}
\nonumber\\
&&\qquad=\sum_{n\in \N^{\Z}\times \N^{\Z}}c(n)\sum_{|j|>j_0}(n_j-n_j')\prod_{\textrm{supp}\,n}q_j^{n_j}\bar q_j^{n_j'}
+O(\eps^A).
\een
Recall from (\ref{2.15},\ref{2.16}) that the monomials in (\ref{5.5}) satisfy
\beno
\Delta(n)\le 20A.
\eeno
So, if $\textrm{supp}\, n\cap \{(-\infty,-j_0)\cup (j_0,+\infty)\}\neq \emptyset$, then
\beno
\textrm{supp}\, n\subset (-\infty,-j_0+20A]\cup [j_0-20A,+\infty),
\eeno
which together with (\ref{5.2}) implies that if $|c(n)|\ge \eps^A$, then
\ben\label{5.6}
\textrm{supp}\, n\subset (-\infty,-j_0-\f N2)\cup (j_0+\f N2,+\infty)\subset (-\infty,-j_0)\cup (j_0,+\infty).
\een
The last set in (\ref{5.6}) is precisely the set that is summed over in (\ref{5.5}). So from (\ref{2.9}), 
$$
\sum_{|j|>j_0}(n_j-n_j')=0
$$
for $n$ with $|c(n)|\ge\epsilon^A$. Thus, only terms where
$|c(n)|<\eps^A$ contribute to (\ref{5.5}) and we get
\ben
\f {d} {dt} \sum_{|j|>j_0}|q'_j(t)|^2< C\eps^A\quad C\,\,\textrm{independent}\,\, \textrm{of} \,\,\epsilon,\label{5.7}
\een
where we used the fact that for the terms with $|c(n)|<\eps^A$
$$
\textrm{supp}\, n\cap \{[-j_0-\f N2,-j_0+\f N2]\cup [j_0-\f N2,j_0+\f N2]\}\neq \emptyset\quad \textrm{and} \quad  \triangle(n), |n|<20 A.
$$
Integrating  (\ref{5.7}) in $t$, we obtain
\ben\label{5.8}
\sum_{|j|>j_0}|q'_j(t)|^2<\sum_{|j|>j_0}|q'_j(0)|^2+C\eps^A t.
\een
Note that the symplectic transformation only acts on a $N$ neighborhood of $\pm j_0$, we obtain
\beno
\sum_{|j|>j_0+N}|q_j(t)|^2=\sum_{|j|>j_0+N}|q_j'(t)|^2<\sum_{|j|>j_0}|q'_j(t)|^2,
\eeno
which together with (\ref{5.8}) gives
\beno
\sum_{|j|>j_0+N}|q_j(t)|^2<\sum_{|j|>j_0}|q_j'(0)|^2+C\eps^A t.
\eeno
On the other hand, the Hamiltonian vector field: 
\beno
\sum\frac{\partial F}{\partial \bar q_j}\frac{\partial}{\partial q_j}- \frac{\partial F}{\partial q_j}\frac{\partial}{\partial \bar q_j}
\eeno
preserves the  $\ell^2$ norm. So we have 
\beno
\sum_{|j|>j_0}|q_j'(0)|^2=\sum_{j\in \Z}|q_j(0)|^2-\sum_{|j|\le j_0}|q_j'(0)|^2<\sum_{|j|>j_0-N}|q_j(0)|^2.
\eeno
Choosing $\bar j_0$ large enough such that
\beno
\sum_{|j|>j_0-N}|q_j(0)|^2<\delta.
\eeno
Then for $t<\f \delta{C}\eps^{-A}$
\beno
\sum_{|j|>j_0+N}|q_j(t)|^2<2\delta
\eeno
with probability
\beno
1-e^{-\f {j_0} {2N}e^{-3N\eps^{\f1{CA}}}}.
\eeno

This completes the proof of Theorem 1.1 after renaming $j_0+N$ as $j_0$ and $2N$ as $N$.


\begin{thebibliography}{100}
\bibitem[AM]{AM} M. Aizenman and S. Molchanov, Localization at large disorder and at extreme energies: an
elementary derivation, Commun. Math. Phys. 157 245-278 (1993).
\bibitem[AF]{AF} C. Albanese and J. Fr\"ohlich, Periodic solutions of some infinite-dimensional Hamiltonian systems associated with non-linear partial difference equations I, Commun. Math. Phys. 116, 475-502 (1988).
\bibitem[AG]{AG} W. Amrein and V. Georgescu, On the characterization of bound states and scattering states in
quantum mechanics, Helv. Phys. Acta 46, 635-658 (1973).
\bibitem[BG]{BG} D. Bambusi and B. Gr\'ebert, Birkhoff normal form for PDE's with tame modulus, Duke Math. J.  135, no. 3, 507-567 (2006).
\bibitem
[BW1]{BW1} J. Bourgain and W.-M. Wang, Diffusion bound for a nonlinear Schr\"{o}dinger equation, 
pp. 21-42 in: 
{\it Mathematical Aspect of Nonlinear Dispersive Equations}, Ann. of Math. Stud., Princeton University press, 
Princeton, NJ,  2007.
\bibitem[BW2]{BW2} J. Bourgain and W.-M. Wang, Quasi-periodic solutions of non-linear random Schršdinger equation, 
J. Eur. Math. Soc. 10, 1-45 (2008).
\bibitem[CFKS]{CFKS} H. L. Cycon, R. G. Froese, W. Kirsch and B. Simon, Schr\"odinger Operators, Springer-Verlag,
1987.
\bibitem[vDK]{vDK} H. von Dreifus and A. Klein, A new proof of localization in the Anderson tight binding model,
Commun. Math. Phys.124, 285-299 (1989).
\bibitem[E]{E} V. Enss,  Asymptotic completeness for quantum-mechanical potential scattering, I. Short-range potentials,
Commun. Math. Phys. 61, 285-291 (1978). 
\bibitem[FS]{FS} J. Fr\"ohlich and T. Spencer, Absence of diffusion in the Anderson tight binding model
for large disorder or low energy, Commun. Math. Phys. 88, 151-184 (1983).
\bibitem[GMP]{GMP} Ya. Gol'dsheid, S. Molchanov and L. Pastur, Pure point spectrum of stochastic one dimensional Schr\"odinger operators, Func. Anal. Appl. 11, 1 (1977).
\bibitem [R]{R} D. Ruelle, A remark on bound states in potential scattering states, Nuovo Cimento 61A, 655-662 (1962).


\end{thebibliography}
\end{document}